\documentclass[12pt]{iopart}

\expandafter\let\csname equation*\endcsname\relax
\expandafter\let\csname endequation*\endcsname\relax

\usepackage{graphicx}
\usepackage{color}
\usepackage{amsmath}
\usepackage{amssymb}
\usepackage{mathrsfs}

\begin{document} 

\title[Ageing first passage times]{Ageing first passage time density in
continuous time random walks and quenched energy landscapes}

\author{Henning Kr{\"u}semann,$^\dagger$ Alja\v{z} Godec$^{\dagger,\ddagger}$ and
Ralf Metzler$^{\dagger,\sharp}$}
\address{$\dagger$ Institute of Physics \& Astronomy, University of Potsdam, D-14476
Potsdam-Golm, Germany\\
$\ddagger$ National Institute of Chemistry, SLO-1000 Ljubljana, Slovenia\\
$\sharp$ Department of Physics, Tampere University of Technology, FI-33101 Tampere,
Finland}

\date{\today}

\begin{abstract}
We study the first passage dynamics of an ageing stochastic process in the
continuous time random walk (CTRW) framework. In such CTRW processes the test
particle performs a random walk, in which successive steps are separated by
random waiting times distributed in terms of the waiting time probability
density function $\psi(t) \simeq t^{-1-\alpha}$ ($0\le \alpha \le 2$). An
ageing stochastic process is defined by the explicit dependence of its
dynamic quantities on the ageing time $t_a$, the time elapsed between its
preparation and the start of the observation. Subdiffusive ageing CTRWs
describe systems such as charge carriers in amorphous semiconductors,
tracer dispersion in geological and biological systems, or the dynamics of
blinking quantum dots.  We derive the exact forms of the first passage time
density for an ageing subdiffusive CTRW in the semi-infinite, confined, and
biased case, finding different scaling regimes for weakly, intermediately,
and strongly aged systems: these regimes, with different scaling laws,
are also found when the scaling exponent is in the range $1<\alpha<2$, for
sufficiently long $t_a$. We compare our results with the ageing motion of a
test particle in a quenched energy landscape. We test our theoretical results
in the quenched landscape against simulations: only when the bias is strong
enough, the correlations from returning to previously visited sites become
insignificant and the results approach the aging CTRW results. With small
or without bias, the ageing effects disappear and a change in the exponent
compared to the case of a completely annealed landscape can be found,
reflecting the build-up of correlations in the quenched landscape.
\end{abstract}

\section{Introduction}

Anomalous diffusion is characterised by the non-linear time dependence
of the mean squared displacement (MSD) of the underlying stochastic
process \cite{metzler2000guide,metzler2004restaur,bouchaud1990anomalous},
and it is widespread in the microscopic world. It can be detected in
numerous systems, ranging form moving charge carriers in amorphous
semiconductors \cite{Scher1975anomalous,steyrleuthner2010transport},
geological \cite{Scher2002geostuff,berkowitz2006geostuff} and biological
\cite{tabei2013biostuff,xu2011surface,wong2004anomalous,kepten2011ergodicity,
burnecki2012universal,tejedor2010quantitative,hoefling2013anomalous}
systems, and similar features also occur in the dynamics of blinking
quantum dots \cite{brokmann2003quantum} and in laser cooling
\cite{schaufler1999keyhole}.

\begin{figure}
 \centering
 \includegraphics[width=.8\textwidth]{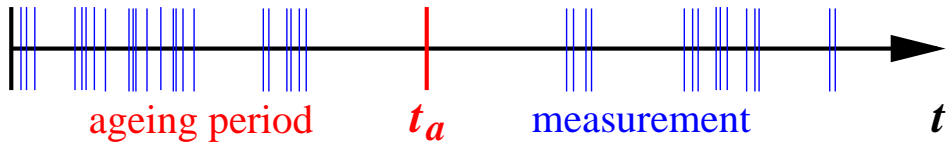}
 \caption{\label{ageing}Sketch of an ageing subdiffusive process. The observation is started after an initial ageing period of length $t_a$, meaning that the origin of the observation time $t$ lies in the point $t_a$. When the scaling exponent is in the range $0<\alpha <1$ statistically longer and longer waiting times appear in the evolution of the process. In this non-stationary scenario physical observables explicitly depend on $t_a$.}
\end{figure}

There are several models describing subdiffusion \cite{metzler2000guide,
bouchaud1990anomalous,metzler2014anomalous,barkai2012strange,sokolov2012models}
and even ultraslow logarithmic diffusion \cite{drager2000strong,sinai1983limiting,
godec2014localisation,sanders2014severe,lomholt2013microscopic}.
The continuous time random walk (CTRW) is one of the best studied anomalous
diffusion processes \cite{metzler2000guide,metzler2004restaur}. It exhibits
subdiffusive behaviour
\begin{equation}
\left<x^2(t)\right> \sim K_{\alpha}t^{\alpha},
\end{equation}
with $0<\alpha<1$ for power law waiting time distributions
\begin{equation}\label{pwrlw}
\psi(\tau) \sim \tau^{-1-\alpha},
\end{equation}
of the waiting times $\tau$ elapsing between any two jumps. Successive waiting times are statistically independent of each other. For $0<\alpha<1$ the distribution $\psi(\tau)$ is characterised by an infinite mean $\left<\tau\right>$. Below we will also consider the case when $1<\alpha<2$ and thus $\left<\tau\right><\infty$, yet the fluctuations around $\left<\tau\right>$ diverge. Theoretical dynamic models that can be effectively analysed within the CTRW theory include dynamic maps \cite{radons1999maps,barkai2003aging2} and particles moving in random (or quenched) energy landscapes \cite{bouchaud1990anomalous,bertin2003subdiffusion,monthus1996models}. CTRWs exhibit several interesting features, such as (i) weak ergodicity breaking, the violation of the Boltzmann-Khinchin equality of the ensemble and lag time averages of physical observables \cite{barkai2012strange,bel2005weak,bouchaud1992weak,lomholt2007subdiffusion,he2008random,burov2010aging,esposito2010relation,jeon2010analysis}, (ii) population splitting, the existence of a growing immobile fraction of particles versus mobile particles \cite{schulz2013aging,schulz2014aging}, and (iii) ageing, the dependence of the time evolution of statistical properties on the time lag $t_a$ between preparation and start of the observation of the system \cite{barkai2003aging2,bouchaud1992weak,schulz2014aging,barkai2003aging}, see figure \ref{ageing}.
Even more significantly, the scaling of physical observables with time changes with the age, as can be seen, for instance, in the MSD, which turns from subdiffusive growth for weak ageing $t_a\ll t$ to seemingly normal (linear) growth for strong ageing $t_a\gg t$ \cite{schulz2014aging,barkai2003aging},
\begin{equation}
\left<x^2(t_a,t)\right> \sim \left\{
\begin{array}{ll}
  K_{\alpha}t^{\alpha},&t_a \ll t\\[0.2cm]
  K_{\alpha}t_a^{\alpha-1} t,&t_a \gg t
 \end{array}
 \right. .
\end{equation}
Some ageing effects persist also for waiting times distributed according to equation (\ref{pwrlw}) with $1 < \alpha <2$ \cite{godreche2001stat,allegrini2002diverging}. Evidence of ageing was found experimentally in biological systems \cite{tabei2013biostuff,weigel2011bioage} as well as in amorphous semiconductors \cite{schubert2013mobility}, glasses \cite{bouchaud1992weak}, and in the blinking dynamics of quantum dots \cite{brokmann2003quantum}. Note that ageing and ergodicity breaking also occurs in other stochastic processes such as Sinai diffusion \cite{godec2014localisation}, scaled Brownian motion \cite{jeon2014scaled,bodrova2015nonergodic}, processes with spatially heterogenous diffusivities \cite{cherstvy2013anomalous,cherstvy2014ageing}, as well as their combinations (generalised diffusion processes \cite{cherstvy2015ageing}) and many-particle systems \cite{sanders2014severe}.

The first passage time, i. e., the time a process needs to reach or cross a certain threshold, is an important tool in the statistical analysis of stochastic processes. The first passage describes, for example, the arrival of a particle at an absorbing barrier or to the binding of a protein to its binding site. The scope of this paper is to describe the effect of aging on the first passage time density (FPTD) for a one dimensional CTRW process in a semi-infinite domain (one absorbing boundary) and in a finite domain (the particle starts between two absorbing boundaries), thereby extending and completing our previous work \cite{krusemann2014first}. In section \ref{nobiassec}, the FPTD for the unbiased cases is derived. The cases of $0<\alpha<1$ and $1<\alpha<2$ are considered separately. In section \ref{biassec} we continue our work on the FPTD for biased diffusion \cite{krusemann2014first} and section \ref{ageqel} is devoted to the ageing of the FPTD in annealed and quenched random energy landscapes. The paper ends with a discussion of the relevance of ageing in first passage processes.

\section{First passage time density for ageing CTRW in semi-infinite and finite domains}\label{nobiassec}
The general idea for the derivation of the first passage time density (FPTD) \cite{redner2001first} is to use the survival probability of the random walker \cite{metzler2000boundary} in analogy to the non-aging case \cite{barkai2001fractional}. The survival probability $\mathscr{S}(t)$ is given by the integral of the probability density function $P_D(x,t_a,t)$ to find the particle at position $x$ at the time $t$ after the ageing period $t_a$ over the spatial domain $D$, in which the walker is confined to move,
\begin{equation}\label{sp}
 \mathscr{S}(s,u) = \int\limits_D{P_D(x,s,u)dx}.
\end{equation}
For simplicity in the following we use the double Laplace transform $(t_a,t \rightarrow s,u)$ of the quantities of interest, i.e.
\begin{equation}
P_D\left(x,s,u\right)=\int\limits^{\infty}_0{\int\limits^{\infty}_0{e^{-ut-st_a}P_D\left(x,t_a, t\right)dtdt_a}}.
\end{equation}
The FPTD $\wp(t_a,t)$ is then given by the negative time derivative of $\mathscr{S}(t_a,t)$, or in double Laplace space as
\begin{equation}\label{afptd}
 \wp(s,u) = \frac{1}{s}-u\mathscr{S}(s,u) .
\end{equation}
The propagator for the unconfined aging CTRW with initial condition $P\left(x,t_a,0\right)=\delta(x)$ is well known. In double Laplace representation it reads \cite{barkai2003aging}
\begin{equation}\label{amw}
 P(x,s,u) = \Phi(s,u)\delta(x) + h(s,u)P_{MW}(x,u).
\end{equation}
The time $t$ we have to wait for the first jump to occur is statistically drawn from the distribution
\begin{equation}\label{hsu}
 h(s,u) = \frac{1}{1-\psi(s)}\frac{\psi(s)-\psi(u)}{u-s}
\end{equation}
of the forward waiting (or recurrence) time, whose statistic is different from that of $\psi(\tau)$ \cite{godreche2001stat}. In equation (\ref{amw}) the PDF $h\left(t_a,t\right)$ is convoluted with the Montroll-Weiss propagator $P_{MW}(x,t)$ \cite{montroll1965random} for the non-ageing CTRW. The first term on the right hand side of equation (\ref{amw}) is the probability of not having made a jump between $t_a$ and $t$,
\begin{equation}\label{P0is}
 \Phi(s,u) = \frac{1}{su} - \frac{h(s,u)}{u}.
\end{equation}
The probability density functions in the semi-infinite domain (with and without force) and in the finite domain all have the following general form
\begin{equation}\label{boundP}
 P_D(x,s,u) = h(s,u)P_D(x,u) + \Phi(s,u) \delta(x).
\end{equation}
Here $P_D(x,u)$ is the known non-ageing solution in the respective cases \cite{metzler2000boundary,barkai2001fractional}. The combination of equations (\ref{sp}), (\ref{afptd}), (\ref{P0is}) and (\ref{boundP}) yields a general expression for the ageing FPTD,
\begin{eqnarray}
 \wp(s,u) &=&\frac{1}{s} - u\Phi(s,u) - uh(s,u)\mathscr{S}(u) \nonumber \\
 &=&h(s,u) \left[1-u\mathscr{S}(u)\right] = h(s,u) \wp(u) \label{afptd2}
\end{eqnarray}
This result implies, that the aging FPTD is the convolution (the Laplace space product) of the forward waiting time density $h\left(t_a,t\right)$ and the non-aging FPTD.
\subsection{Semi-infinite domain}
\begin{figure}
 \centering
 \includegraphics[width=.8\textwidth]{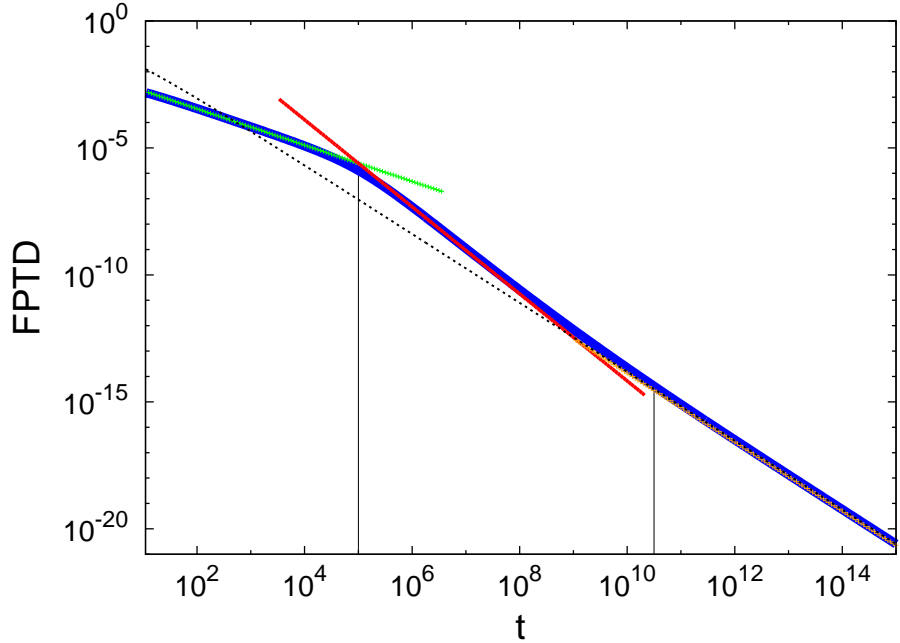}
 \caption{FPTD in the semi-infinite case (blue line). One can see the three scaling regimes of equation (\ref{3scale}), which are separated by $t_a = 10^5$ and $t^* \approx 3.15 \times 10^{10}$ indicated by the vertical lines. The parameters are $\alpha = 5/7$, $x_0=1$ and $K_{\alpha} = 1/5$, green: $\sim t^{-\alpha}$ ($t \ll t_a$), red: $\sim t^{-1-\alpha}$ ($t_a \ll t \ll t^*$), orange: $\sim t^{-1-\alpha /2}$ ($t \gg t^*$) \label{3reg}. The dotted line is the non-ageing FPTD equation (\ref{levystab}).}
\end{figure}

\begin{figure}
 \centering
 \includegraphics[width=.8\textwidth]{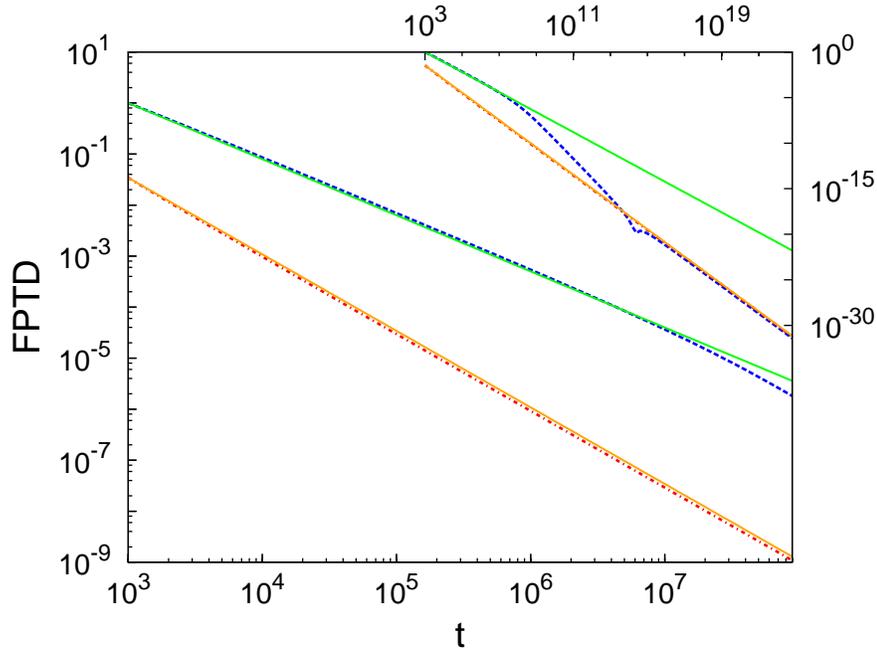}
 \caption{Numerical Laplace inversion of equation (\ref{hsu}) multiplied with the Brownian FPTD $\wp(u) = \exp{\left(-\frac{a}{\sqrt{K_{\alpha}}} \sqrt{u}\right)}$ for $\alpha=1.1$ (blue, dashed line, top) and $\alpha=1.9$ (red, dashed-dotted line, bottom) in the strong ageing limit (the Laplace variable is chosen as $s=10^{-8}$, which roughly corresponds to an ageing time $t_a\approx 10^8$). The first case with $\alpha=1.1$ is compared to the forward waiting time (\ref{FWT2}) (green), the second case with $\alpha=1.9$ to the Brownian FPTD (orange), see the top right corner: same plots, larger scale: all three regimes (cf. equation (\ref{ag13regimes})) are shown, the kink is due to numerical errors in the Laplace inversion. The function plotted is $\wp(t,s)$ with fixed $s$, which behaves similar to $\wp(t,t_a)$, since in the first and in the final time regime the value does not depend on the second variable. We only have one less numerical inverse Laplace transformation and thus one less source of numerical errors.\label{ag1}}
\end{figure}

The FPTD of a non-aged CTRW on the semi-axis, starting at $x=0$ and with the absorbing boundary located at $x=x_0$, is proportional to a one-sided L{\'e}vy stable law (L{\'e}vy-Smirnov law) \cite{hughes1995random}
\begin{equation}\label{levystab}
 \wp(t) = \left(\frac{K_{\alpha}}{x_0^2}\right)^{\frac{1}{\alpha}} \ell_{\frac{\alpha}{2}}\left(\left[\frac{x_0^2}{K_{\alpha}}\right]^{\frac{1}{\alpha}}t\right),
\end{equation}
which is defined via its Laplace transform in the form
\begin{equation}
 \wp(u) = \exp{\left(-\frac{x_0}{\sqrt{K_{\alpha}}}u^{\alpha/2}\right)}.
\end{equation}
The forward waiting time distribution has different expressions for the two ranges of $\alpha$ we consider, respectively. For $0<\alpha <1$ one finds the form \cite{schulz2014aging,godreche2001stat}
\begin{equation}\label{htat}
 h(t_a,t) = \frac{\sin{(\alpha \pi)}}{\pi} \left(\frac{t_a}{t}\right)^{\alpha} \frac{1}{t_a + t}
\end{equation}
which is valid for all ageing times. For $1<\alpha<2$ the mean waiting time is finite ($\left< \tau \right> <\infty$) but the fluctuations are unbound ($\left< \tau^2 \right> \rightarrow \infty$) \cite{shlesinger1974anomalous}. In this case the asymptotic forms are known, which are $h(t) = 1$ for $t_a \rightarrow 0$, while for $t_a \rightarrow \infty$ \cite{schulz2014aging,godreche2001stat}
\begin{equation}\label{FWT2}
 h(t) = \frac{1}{\left<t\right>} \left(\frac{\tau}{t}\right)^{\alpha}.
\end{equation}
This is not sufficient to find all regimes of the FPTD. We use the full form (\ref{hsu}) in the Laplace domain instead, in combination with the Laplace space waiting time density \cite{godreche2001stat}
\begin{equation}\label{wtag1}
 \psi(u) \simeq 1 + \left<t\right> u + \tau^{\alpha}\Gamma[1-\alpha]u^{\alpha},
\end{equation}
valid in the limit $u\rightarrow 0$. For $\alpha>2$, ageing effects are formally present, but decay too fast to be detectable \cite{schulz2014aging}, compare similar transient ageing and weak ergodicity breaking effects in asymptotically ergodic systems \cite{kursawe2013transient,jeon2012inequivalence,jeon2013anomalous}.
We first deal with the case $0<\alpha<1$ and come back to the second case below. The central result for the aging FPTD is the Laplace convolution in $t$ \cite{krusemann2014first} \footnote{I.e., $f(t)\otimes g(t)$ is defined as $\int\limits^t_0{g(\tau)f(t-\tau)d\tau}$.}
\begin{equation}
 \wp(t_a,t) = \frac{\sin{(\alpha \pi)}}{\pi} \left(\frac{t_a}{t}\right)^{\alpha} \frac{1}{t_a + t} \otimes \left(\frac{K_{\alpha}}{x_0^2}\right)^{\frac{1}{\alpha}} \ell_{\frac{\alpha}{2}}\left(\left[\frac{x_0^2}{K_{\alpha}}\right]^{\frac{1}{\alpha}}t\right),
\end{equation}
which follows directly from equations (\ref{afptd2}), (\ref{levystab}), (\ref{htat}) and the convolution theorem of the Laplace transformation \cite{davies2002integral}.
For practical reasons one can derive a result in form of a series expansion of equation (\ref{afptd2}) in powers of $t$, using the single Laplace transform of $h(t_a,t)$ with respect to $t$,
\begin{equation}
 h(t_a,u) = \exp{(t_au)} \Gamma\left(\alpha,t_a u\right) \Gamma^{-1}(\alpha)
\end{equation}
A closed form solution of the inverse of the full Laplace space result cannot be found. The incomplete $\Gamma$-function can be expanded in a series
\begin{equation}
 \Gamma \left(\alpha,t_au\right) = \Gamma(\alpha)\left[1 - \exp{(-t_au)}\sum\limits^{\infty}_{n=0}{\frac{\left(t_au\right)^{n+\alpha}}{\Gamma(1+\alpha+n)}}\right] \label{inclGamma}.
 \end{equation}
Similarly, the series for the non-aging FPTD is
 \begin{equation}
 \wp(u) = \sum\limits^{\infty}_{m=0}{\left(-\frac{x_0}{\sqrt{K_{\alpha}}}\right)^m u^{m \alpha/2}}.
\end{equation}
With these series, equation (\ref{afptd2}) turns into
\begin{eqnarray}
 \wp(t_a,u) &=& \exp{(t_au)}\sum\limits^{\infty}_{m=0}{\left(-\frac{x_0}{\sqrt{K_{\alpha}}}\right)^m u^{m \alpha/2}} \nonumber \\
 &&- \sum\limits_{m,n=0}^{\infty}{\frac{t_a^{n+\alpha}}{m!\Gamma(1+n+\alpha)} \left(-\frac{x_0}{\sqrt{K_{\alpha}}}\right)^m u^{n+\alpha+m \alpha /2}}.
\end{eqnarray}
The inverse Laplace transform of this expression can be obtained term-by-term. The exponential function yields a shifted Laplace transform $u \rightarrow t+t_a$ of the remainder and the powers of $u$ are inverted using the Tauberian theorem \cite{feller1971probability}
\begin{equation}
 \mathscr{L}^{-1}_t\left\{u^b\right\} = \frac{t^{-1-b}}{\Gamma(-b)}.
\end{equation}
This formula for the inverse Laplace transform $\mathscr{L}^{-1}$ is valid for $u \rightarrow 0$ and $t \rightarrow \infty$, and thus we find the aging FPTD
\begin{eqnarray}
 \wp(t_a,t) &=& \delta(t_a+t) + \sum\limits_{m=1}^{\infty}{\left(-\frac{x_0}{\sqrt{K_{\alpha}}}\right)^m\frac{(t+t_a)^{-1-m \alpha /2}}{m! \Gamma\left(-m \alpha /2\right)}} \nonumber  \\
 &&\hspace*{-0.5cm}-\sum\limits_{n,m=0}^{\infty}{\frac{t_a^{n+\alpha}}{m! \Gamma(1+n+\alpha)}\left(-\frac{x_0}{\sqrt{K_{\alpha}}}\right)^m \frac{t^{-1-n-\alpha(1+m/2)}}{\Gamma\left[-n-\alpha(1+m/2)\right]}}.\label{fptsemi}
\end{eqnarray}
The asymptotic behaviour for $t\rightarrow \infty$, meaning the dominating powers of $t$, are found from comparison of the terms with the smallest exponents in $t$, assuming that we are in the weak ageing regime $t\gg t_a$. We find two different regimes with a crossover at 
\begin{equation}
t^* = t_a^2\left[2\frac{\sqrt{K_{\alpha}} \Gamma\left(1-\alpha /2\right)\sin{(\pi\alpha)}}{x_0\pi\alpha}\right]^{\frac{2}{\alpha}}.
\end{equation}
At $t^*$, the ageing effects become negligible and the final non-ageing regime is reached. Note that the crossover time $t^*$ depends on the initial condition $x_0$.
In the strong ageing limit $t_a\gg t$ one has to employ the asymptotic expansion of the incomplete $\Gamma$-function for $ut_a \rightarrow \infty$, which is found using the integral definition and partial integration --- this result cannot be found from the Taylor series. The dominating term is:
\begin{equation}
 \Gamma\left(\alpha,t_au\right) \approx \left(ut_a\right)^{\alpha-1} \exp{\left(-ut_a\right)},
\end{equation}
for $ut_a \rightarrow \infty$. In this limit the forward waiting time is dominating the rest of the first passage process, and consequently, the inverse Laplace transform of this asymptotic FPTD yields to lowest order
\begin{equation}\label{ltl1}
 \wp(t_a,t) = \frac{\sin{(\pi\alpha)}}{\pi}t_a^{\alpha -1} t^{-\alpha}.
\end{equation}
This is the just the limit of $h(t_a,t)$ for $t_a\gg t$. We have thus identified three regimes, separated by the crossover times $t_a$ and $t^*$, namely
\begin{equation}\label{3scale}
 \wp(t_a,t)\sim \left\{
 \begin{array}{ll}
  \displaystyle \frac{\sin{(\pi\alpha)}}{\pi}t_a^{\alpha -1} t^{-\alpha},&t_a \gg t\\[0.3cm]
  \displaystyle \frac{\sin{(\pi\alpha)}}{\pi}t_a^{\alpha} t^{-1-\alpha},&t_a \ll t \ll t^*\\[0.3cm]
  \displaystyle \frac{x_0\alpha}{2\Gamma(1-\alpha /2)\sqrt{K_{\alpha}}}t^{-1-\alpha /2},&t^* \ll t
 \end{array}
 \right.
\end{equation}
The first exponent of $t$ is $-\alpha$, the steepest exponent $-1-\alpha$ can be found in the intermediate time regime. In the asymptotic regime one finds the intermediate exponent $-1-\alpha/2$ equivalent to the non-ageing case \cite{metzler2004restaur,metzler2000boundary}. Since the crossover time $t^*$ decreases with the initial distance $x_0$ of the diffusing particle from the absorbing boundary, we find that the final regime takes over faster for larger $x_0$. As a result, the probability density decays slower than for short distances, which is intuitively expected.
The three regimes (\ref{3scale}) can be clearly distinguished in simulations, as shown in figure \ref{3reg}. One can see that the non-ageing power law is only reached after extremely long times. On short or intermediate time scales, the ageing effects are dominant and if the corresponding parameters allow both power laws persist long enough to be in principle detectable in experiments.
\newline In the case of $1<\alpha<2$, for which the mean waiting time $\left<\tau\right>$ is finite, one finds the long time limit of the FPTD in the non-ageing case to be the same Sparre-Andersen law as for Brownian diffusion (figure \ref{ag1}),
\begin{equation}\label{ltlbrown}
 \wp(t) \sim \frac{x_0}{2\sqrt{K\pi}}t^{-3/2}.
\end{equation}
For the ageing case, one uses the general form (\ref{hsu}) of the forward waiting time and the waiting time density (\ref{wtag1}). The ageing FPTD (cf. equation (\ref{afptd2})) is analysed in a similar way as for $0<\alpha<1$. As a first result, three power laws can be distinguished for $1<\alpha<3/2$, which are separated by the crossover times $t_a$ and
\begin{equation}
 t^*=\left[\frac{\tau^{\alpha}\sqrt{\pi K}}{2\alpha x \left<t\right>}\right]^{2/(2\alpha -1)},
\end{equation}
which is again the time that separates the ageing regimes from the non-ageing one.
The power law behaviour in those regimes is, respectively,
\begin{equation}\label{ag13regimes}
 \wp(t_a,t)\sim \left\{
 \begin{array}{ll}
  \displaystyle \frac{1}{\left<t\right>}\left(\frac{\tau}{t}\right)^{\alpha},&t_a \gg t\\[0.4cm]
  \displaystyle \frac{\tau^{\alpha} t_a}{\left<t\right>\alpha}t^{-1-\alpha},&t_a \ll t \ll t^*\\[0.4cm]
  \displaystyle \frac{x_0}{2\sqrt{K\pi}}t^{-3/2},&t^* \ll t
 \end{array}
 \right.
\end{equation}
The first regime is the one, in which the forward waiting time density dominates the FPTD (cf. equation (\ref{FWT2})). The last power law is that of the Brownian limit. In the time range $t \gg t^*$ the system has evolved for so long that the fluctuations of the aged waiting times around $\left<\tau\right>$ become negligible. Only in the crossover regime the dominating term depends on the ageing time explicitly. The three regimes are shown in figure \ref{ag1}, where the middle regime is not indicated by a power law line as the picture would be overloaded.
For $\alpha>3/2$, these regimes collapse to a single non-ageing regime (cf. figure \ref{ag1}), which is just the Brownian one (equation (\ref{ltlbrown})).
We will see in section \ref{biassec}, that the introduction of a constant bias will change the long time behaviour severely.
\subsection{Bounded domain}\label{boxsec}
In a bounded domain $[-a,a]$, we find the solution of the non-ageing FPTD using a trigonometric basis for expansion that vanishes at the boundaries $x=\pm a$. This is equivalent to an infinite application of the method of images \cite{feller1971probability}. For the symmetric initial condition $P(x,t_a,t=0) = \delta(x)$, the solution reads
\begin{equation}\label{propsymm}
 P_D(x,s,u) = \frac{1}{a} \sum\limits^{\infty}_{m=0}{\mathrm{Re}\left\{\exp{\left(i\frac{(2m+1)\pi}{2a}x\right)}\right\}P\left(k=\frac{(2m+1)\pi}{2a},s,u\right)},
\end{equation}
where $\mathrm{Re}\left\{\cdot\right\}$ is the real part of the argument. To solve the asymmetric initial condition $P(x,t_a,t=0)=\delta(x-x_0)$, $ -a\le x_0 \le a$ we shift the sum to the coordinate system $x^{\prime} = x+x_0$ and introduce the factors $a_m$, so that
\begin{equation}
 P_D(x^{\prime},s,u) = \frac{1}{a} \sum\limits^{\infty}_{m=0}{\mathrm{Re}\left\{a_m\exp{\left(i\frac{(2m+1)\pi}{2a}(x^{\prime}-x_0)\right)}\right\}P\left(k=\frac{(2m+1)\pi}{2a},s,u\right)}\label{asymprop}.
\end{equation}
This solution is in fact the propagator for the symmetric initial condition $P(x,t_a,t=0)=\frac{1}{2}\left[\delta(x-x_0) + \delta(x+x_0)\right]$, but the resulting FPTD is the same as for the asymmetric case. A way to find the propagator for the asymmetric initial condition is mentioned in Ref. \cite{redner2001first}. With equation (\ref{asymprop}), we have to match the $\delta$ initial condition (at $x^{\prime}=0$, due to the coordinate shift) and thus we identify
\begin{equation}
 a_m = \exp{\left(i\frac{(2m+1)\pi}{2a}x_0\right)}.
\end{equation}
The result in the original coordinates is now
\begin{equation}
 P_D(x,s,u) = \frac{1}{a} \sum\limits^{\infty}_{m=0}{\mathrm{Re}\left\{\exp{\left(i\frac{(2m+1)\pi}{2a}(x+x_0)\right)}\right\}P\left(k=\frac{(2m+1)\pi}{2a},s,u\right)}.
\end{equation}
We replace the ageing unconfined propagator using equation (\ref{amw})
\begin{eqnarray}
\hspace*{-2.5cm} P_D(x,s,u) &\hspace*{-0.5cm}=&   \hspace*{-0.05cm}\frac{h(s,u)}{a}\sum\limits^{\infty}_{m=0}{\mathrm{Re}\left\{\exp{\left(i\frac{(2m+1)\pi}{2a}(x-x_0)\right)}\right\}P_{MW}\left(k=\frac{(2m+1)\pi}{2a},u\right)} \nonumber \\
 &&+ \frac{\Phi(s,u)}{a}\sum\limits^{\infty}_{m=0}{\mathrm{Re}\left\{\exp{\left(i\frac{(2m+1)\pi}{2a}(x-x_0)\right)}\right\}}.
\end{eqnarray}
In the first line, we just have the product $h(s,u)P_D(x,u)$ on the right hand side. The sum of the exponentials can be transformed into a sum of $\delta$-functions, using the Poisson summation formula, and
\begin{equation}
\sum\limits^{\infty}_{n=0}{\exp{(i\pi n x)}} = \sum\limits^{-1}_{n=-\infty}{\exp{(i \pi n x)}}.
\end{equation}
We find
\begin{eqnarray}
 \hspace*{-2.0cm}\sum\limits^{\infty}_{m=0}{\mathrm{Re}\left\{\exp{\left(i\frac{(2m+1)\pi}{2a}(x-x_0)\right)}\right\}} &=& \frac{1}{2} \sum\limits^{\infty}_{m=-\infty}{\mathrm{Re}\left\{\exp{\left(i\frac{(2m+1)\pi}{2a}(x-x_0)\right)}\right\}} \nonumber \\
 &=&a \sum\limits^{\infty}_{m=-\infty}{\delta(x-x_0-2a)}.
\end{eqnarray}
Only the peak for $m=0$ is within the boundaries of the domain and thus $P_D(x,s,u)$ has the form of equation (\ref{boundP}), with a $\delta$ initial condition at $x=x_0$ instead of $x=0$. But since the integral for the survival probability is taken over the whole domain, this does in fact not make a difference for the FPTD. In Laplace space, we obtain \cite{metzler2000boundary}
\begin{eqnarray}
\hspace*{-0.7cm} \wp(t_a,u) &=& \Gamma\left(\alpha,t_au\right)\exp{(t_au)} \Gamma^{-1}(\alpha)  \nonumber \\
 &&\hspace*{-0.8cm}\times \left(1-\sum\limits^{\infty}_{m=0}{\mathrm{Re}\left\{\frac{4 (-1)^m \exp{\left(i\frac{(2m+1)\pi}{2a}x_0\right)}}{(2m+1)\pi}\right\} \frac{u^{\alpha}}{u^{\alpha}+K_{\alpha}\frac{(2m+1)^2 \pi^2}{4a^2}}}\right) .\label{fullsolbound}
\end{eqnarray}
To find the solution of the non-ageing part in the form of a power series, we expand the last fraction in equation (\ref{fullsolbound}) about $u=0$,
\begin{eqnarray}
 \frac{u^{\alpha}}{u^{\alpha}+K_{\alpha}\frac{(2m+1)^2 \pi^2}{4a^2}} &=& \frac{4a^2 u^{\alpha}}{K_{\alpha}(2m+1)^2\pi^2} \sum\limits^{\infty}_{n=0}{(-1)^n\left[\frac{4a^2 u^{\alpha}}{K_{\alpha}(2m+1)^2\pi^2}\right]^n} \nonumber \\
 &=& - \sum\limits^{\infty}_{n=0}{\left[-\frac{4a^2 u^{\alpha}}{K_{\alpha}(2m+1)^2\pi^2}\right]^{n+1}} \nonumber \\
 &=& -\sum\limits^{\infty}_{n=1}{\left[-\frac{4a^2 u^{\alpha}}{K_{\alpha}(2m+1)^2\pi^2}\right]^n}.
\end{eqnarray}
Changing the order of summation, we get
\begin{eqnarray}\label{FPTDconf}
 \wp(t_a,u) &=& \Gamma\left(\alpha,t_au\right)\exp{(t_au)} \Gamma^{-1}(\alpha) \\
 &&\hspace*{-1.5cm}\times \left(1+  \sum\limits^{\infty}_{n=1}{\left[-\frac{a^2 u^{\alpha}}{K_{\alpha}}\right]^n \sum\limits^{\infty}_{m=0}{\mathrm{Re}\left\{\frac{2^{2n+2} (-1)^m \exp{\left(i\frac{(2m+1)\pi}{2a}x_0\right)}}{(2m+1)^{2n+1}\pi^{2n+1}}\right\}}}\right). \nonumber 
\end{eqnarray}
As the series in $m$ equals unity for $n=0$,
\begin{eqnarray}
 \sum\limits^{\infty}_{m=0}{\mathrm{Re}\left\{\frac{4 (-1)^m \exp{\left(i\frac{(2m+1)\pi}{2a}x_0\right)}}{(2m+1)\pi}\right\}} &=& \frac{4}{\pi}\mathrm{Re}\left\{\arctan{\left(i \pi \frac{x_0}{2 a}\right)}\right\} \nonumber \\
 &=& 1,
\end{eqnarray}
for $-a \le x_0 \le a$, we can write the $1$ in equation (\ref{FPTDconf}) as the zeroth element of the series in $n$. Furthermore, the series in $m$ can be written as a Lerch function $\mathfrak{K}$ \cite{bateman1953higher}:
\begin{eqnarray}
 &&\sum\limits^{\infty}_{m=0}{\mathrm{Re}\left\{\frac{2^{2n+2}(-1)^m \exp{\left(i\frac{(2m+1)\pi}{2a}x_0\right)}}{(2m+1)^{2n+1}\pi^{2n+1}}\right\}} \nonumber \\
 &&= \frac{2}{\pi^{2n+1}}\mathrm{Re}\left\{\exp{\left(i \pi \frac{x_0}{2a}\right)} \mathfrak{K}\left(\frac{a+x_0}{2a},2n+1,\frac{1}{2}\right)\right\},
\end{eqnarray}
so that the FPTD as a series in powers of $u^{\alpha}$ reads
\begin{eqnarray}
 \hspace*{-2.0cm}\wp(t_a,u) &\hspace*{-0.65cm}=& \Gamma\left(\alpha,t_au\right) \exp{(t_au)} \Gamma^{-1}(\alpha) \nonumber \\
 &&\hspace*{-0.5cm}\times\sum\limits^{\infty}_{n=0}{u^{n\alpha}\left[-\frac{a^2}{K_{\alpha}}\right]^n \frac{2}{\pi^{2n+1}} \mathrm{Re}\left\{ \exp{\left(i \pi \frac{x_0}{2a}\right)} \mathfrak{K}\left( \frac{a+x_0}{2a},2n+1,\frac{1}{2} \right) \right\}}.
\end{eqnarray}
Using also equation (\ref{inclGamma}), the Laplace inversion $u\rightarrow t$ yields
\begin{eqnarray}
 \hspace*{-1.5cm}\wp(t_a,t) &\hspace*{-0.25cm}=& \wp(t_a+t) - \sum\limits_{n,m}\frac{t_a^{n+\alpha}}{\Gamma(1+n+\alpha)}\left(-\frac{a^2}{K_{\alpha}}\right)^m  \nonumber \\
 &&\hspace*{-0.5cm}\times \frac{t^{-1-n-\alpha(1+m)}}{\Gamma\left[-n-\alpha(1+m)\right]} \frac{2}{\pi^{2n+1}} \mathrm{Re}\left\{\exp{\left(\frac{i \pi x_0}{2a}\right)} \mathfrak{K}\left(\frac{a+x_0}{2a},2n+1,\frac{1}{2}\right) \right\} 
\end{eqnarray}
In the symmetric case ($x_0=0$), one can simplify this due to the relation between the Lerch function and the even Euler numbers \cite{bateman1953higher}
\begin{equation}
 \frac{2}{\pi^{2m+1}}\mathfrak{K}\left(1/2,2m+1,1/2\right)=\frac{\left|E_{2m}\right|}{(2m)!}
\end{equation}
It should be noted, that the symmetric result is more intuitively found when the coefficients are written as an infinite series of cosines instead of Lerch functions:
\begin{eqnarray}
&&\frac{2}{\pi^{2n+1}}\mathrm{Re}\left\{\exp{\left(\frac{i \pi x_0}{2a}\right)} \mathfrak{K}\left(\frac{a+x_0}{2a},2n+1,\frac{1}{2}\right) \right\} \nonumber \\
&&=\frac{2}{\pi^{2n+1}}\sum\limits^{\infty}_{m=0}{\frac{\cos{\left(\pi n + \frac{\pi x_0}{2a}\left[1+2n\right]\right)}}{\left(m+\frac{1}{2}\right)^{2n+1}}}.
\end{eqnarray}
Setting $x_0$ to zero, we are left with the cosines of $n\pi$, which is just $(-1)^n$. This alternating series is exactly the series for the Euler numbers,
\begin{equation}
\left|E_{2n}\right| = \frac{2^{2n+2}(2n)!}{\pi^{2n+1}} \sum\limits^{\infty}_{m=0}{\frac{(-1)^m}{(2m+1)^{2n+1}}}.
\end{equation}
The behaviour of the FPTD for short ageing times is
\begin{eqnarray}
 \hspace*{-1.0cm}\wp(t_a,t) &\sim& t^{-1-\alpha}\left[-\frac{t_a^{\alpha}}{\Gamma(-\alpha)\Gamma(1+\alpha)}\right. \nonumber \\ 
 &&\hspace*{1.0cm}\left. - \frac{2 a^2}{K_{\alpha} \pi^3 \Gamma(-\alpha)} \mathrm{Re}\left\{\exp{\left[\frac{i \pi x_0}{2a}\right]} \mathfrak{K}\left[\frac{a+x_0}{2a},3,1/2\right]\right\}\right],
\end{eqnarray}
or in the symmetric case
\begin{equation}
 \wp(t_a,t) \sim t^{-1-\alpha} \left[- \frac{t_a^{\alpha}}{\Gamma(-\alpha) \Gamma(1+\alpha)} - \frac{a^2}{2K_{\alpha}\Gamma(-\alpha)}\right].
\end{equation}
If $t_a$ is sufficiently large, the second term in the square brackets, and thus also any asymmetry effect, becomes irrelevant.
\begin{figure}
 \centering
 \includegraphics[width=.75\textwidth]{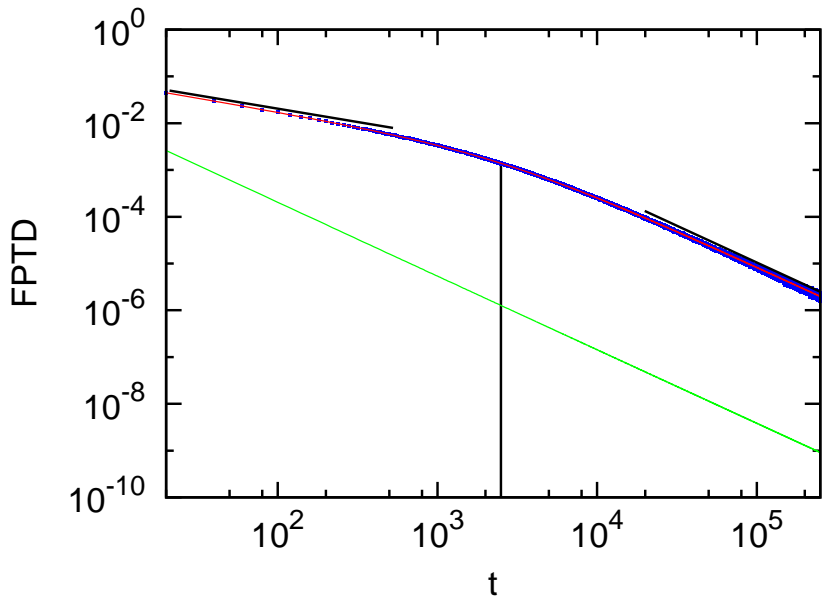}
 \caption{Simulation (blue squares) and exact FPTD (red line) of an ageing CTRW in a box (green (lower) line: result without ageing); the power laws ($\sim t^{-\alpha}$ for $t \ll t_a$ and $\sim t^{-1-\alpha}$ for $t \gg t_a$) are indicated by black lines, the crossover at $t_a$ is indicated by a vertical black line ($t_a = 2.5\times 10^3$, $\alpha = 4/7$, $K_{\alpha}=1/2$, $a=1$, and $\sim 10^8$ runs).\label{symm}}
\end{figure}
In the strong ageing limit $t_a \gg t$ (cf. figure \ref{symm}) the behaviour of the FPTD is dominated again by the forward waiting time, and thus the result is the same as in the semi-infinite case (cf. equation (\ref{ltl1})).
The power laws in the long time limit are thus (cf. Ref. \cite{krusemann2014first})
\begin{equation}
 \wp(t_a,t)\sim \left\{
 \begin{array}{ll}
  \displaystyle \frac{\sin{(\pi\alpha)}}{\pi}t_a^{\alpha -1} t^{-\alpha},&t_a \gg t\\[0.3cm]
  \displaystyle \left[\frac{\sin{(\pi\alpha)}}{\pi}t_a^{\alpha} + \frac{\alpha a^2}{2K_{\alpha}\Gamma(1-\alpha)}\right]t^{-1-\alpha},&t_a \ll t
 \end{array}
 \right. .
\end{equation}
One can see that a nonzero ageing time will yield an offset in the limit $t_a\ll t$, and different from the semi-infinite case, the non-ageing and ageing curves do not collapse onto the same curve (cf. figures \ref{symm} and \ref{3reg}).
In the case $1<\alpha<2$, the non-ageing FPTD is approximately an exponential distribution, which is found in the case of normal diffusion (similar to the results derived in \cite{margolin2000first}), and yields a finite mean first passage time. This means, that, after the first step is made, the particle will be absorbed very fast. The long time behaviour is thus dominated by the forward waiting time density for long ageing times (equation (\ref{FWT2})) and an exponential decay in the case of a weakly aged CTRW.
\section{First passage density for biased ageing CTRW}\label{biassec}
\begin{figure}
 \centering
 \includegraphics[width=.75\textwidth]{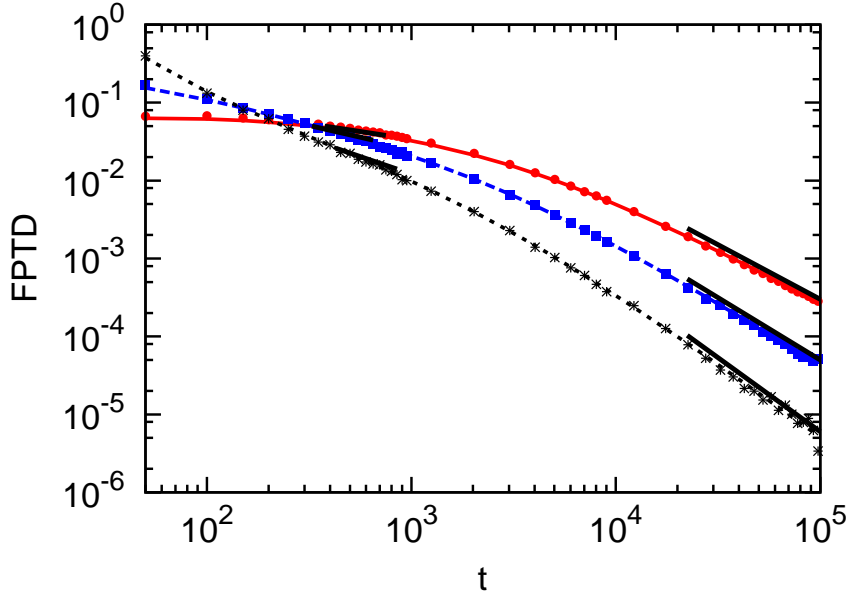}
 \caption{Simulation of the FPTD of a biased ageing CTRW for three different anomalous exponents $\alpha = 0.4$ (red, circles), $\alpha=0.6$ (blue, squares) and $\alpha=0.9$ (black, stars). The parameters are $F=1$, $T=1$, $t_a = 2500$, $x_0=5$ and $K_{\alpha}=1/2$. The lines are the theoretical results (\ref{theoforce}), the short black slopes indicate the power laws (\ref{powerforce}). \label{forcestuff}}
\end{figure}

We now introduce a constant bias $F$ acting on the random walker. This problem is handled using the method of images with corrections due to the external field, or by direct solution of the advection-diffusion equation \cite{redner2001first,barkai2001fractional}. The propagator on the semi axis reads 
\begin{equation}\label{forceprop}
 P_D(x,s,u) = P_F(x,s,u) - \exp{\left(\frac{Fx_0}{T}\right)} P_F(x-2x_0,s,u).
\end{equation}
In this, the value of the Boltzmann constant is set to unity. The function $P_F(x,s,u)$ is the unconfined ageing propagator under a constant bias,
\begin{equation}
 P_F(x,s,u) = \Phi(s,u) \delta(x) + h(s,u) P_F(x,u),
\end{equation}
and $P_F(x,u)$ is its non-ageing equivalent \cite{barkai2001fractional}
\begin{equation}
 P_F(x,u) = \frac{F u^{\alpha-1}\tau^{(*)\alpha}}{T\sqrt{1+ 4 (u\tau^{(*)})^{\alpha}}}\exp{\left(\frac{F\left(x-\sqrt{1+4 (u\tau^{(*)})^{\alpha}}|x|\right)}{2T}\right)}
\end{equation}
with the scaling time $\tau^{(*)\alpha} =T^2/(F^2K_{\alpha})$. From the additive form of equation (\ref{forceprop}) one sees immediately, that equation (\ref{boundP}) remains valid. Using \cite{barkai2001fractional}
\begin{equation}
 \wp(u) = \exp{\left(\frac{Fx_0}{2T}\left[1-\sqrt{1+4(u\tau^{\left(*\right)})^{\alpha}}\right]\right)}
\end{equation}
one consequently finds the ageing FPTD via equation (\ref{afptd2}). We can write the non-ageing result as an infinite convolution of one-sided L{\'e}vy stable laws or as a series in powers of $t$. In the first case, we expand the square root in the exponent using \cite{abramowitz1964handbook}
\begin{equation}\label{expansion}
 (1+x)^{a-1} = \sum\limits^{\infty}_{k=0}{\frac{\Gamma(a)}{\Gamma(a-k)k!} x^k}, \quad \left|x\right| \le 1.
\end{equation}
An exponential to the power of a sum of terms can be written as a product of exponentials, and we thus find
\begin{equation}
 \wp(u) = \prod\limits^{\infty}_{k=1}{\exp{\left(x_0\frac{\Gamma(3/2)}{\Gamma(3/2-k)k!}\left(\frac{1}{K_{\alpha}}\right)^k\left(\frac{2T}{F}\right)^{2k-1} u^{k\alpha}\right)}}.
\end{equation}
The exponential function is the Laplace pair of the one-sided L{\'e}vy stable function according to equation (\ref{levystab}) and thus, the non-ageing FPTD is an infinite convolution of L{\'e}vy stable functions \cite{krusemann2014first}\footnote{We define $\bigotimes^{\infty}_{k=1}f(t)$ as $f(t) \otimes f(t) \otimes f(t)\ldots \otimes f(t)$.}
\begin{equation}
 \wp(t) = \bigotimes^{\infty}_{k=1}\mathcal{D}(\alpha,k)\ell_{k\alpha}\left(\mathcal{D}(\alpha,k)t\right)
\end{equation}
with
\begin{equation}\mathcal{D}(\alpha,k) = \left[x_0\frac{\Gamma(3/2)}{\Gamma(3/2-k)k!}\left(\frac{1}{K_{\alpha}}\right)^k\left(\frac{2T}{F}\right)^{2k-1}\right]^{-1/(k\alpha)}.
\end{equation}
The ageing FPTD is just the convolution with the forward waiting time. For the power series, we first expand the non-ageing FPTD
\begin{equation}
 \wp(u) = \exp{\left(\frac{Fx_0}{2T}\right)}\sum\limits_{n=0}^{\infty}{\left(-\frac{Fx_0}{2T}\right)^n\frac{1}{n!} \left[1+4(u\tau^{(*)})^{\alpha}\right]^{n/2}}.
\end{equation}
Then we use the series expansion (\ref{expansion}) again, only this time with the exponent depending on $n$. Change of order of summations yields
\begin{equation}
\wp(u) = \exp{\left(\frac{Fx_0}{2T}\right)}\sum\limits_{k=0}^{\infty}{\left[4(u\tau^{(*)})^{\alpha}\right]^k\frac{1}{k!}\sum\limits_{n=0}^{\infty}{\left(-\frac{Fx_0}{2T}\right)^n\frac{1}{n!}\frac{\Gamma\left(\frac{n+2}{2}\right)}{\Gamma\left(\frac{n+2}{2}-k\right)}}}
\end{equation}
The sum in $n$ can be turned into a combination of hypergeometric functions. First we separate the even and odd numbered terms into two sums and change the index of summation,
\begin{equation}
 \sum\limits_{n=0}^{\infty}{\frac{\left(\frac{Fx_0}{2T}\right)^n}{n!}\frac{\Gamma\left(\frac{n+2}{2}\right)}{\Gamma\left(\frac{n+2}{2}-k\right)}} = \sum\limits_{m=0}^{\infty}{\frac{\left(\frac{Fx_0}{2T}\right)^{2m}}{\Gamma(2m+1)}\frac{\Gamma\left(m+1\right)}{\Gamma\left(m+1-k\right)}} + \sum\limits^{\infty}_{m=0}{\frac{\left(\frac{Fx_0}{2T}\right)^{2m+1}}{\Gamma(2m+2)}\frac{\Gamma\left(m+\frac{3}{2}\right)}{\Gamma\left(m+\frac{3}{2}-k\right)}}.
\end{equation}
Then, using the duplication formula
\begin{equation}
\frac{\Gamma\left(z+\frac{1}{2}\right)}{\Gamma\left(2z\right)} = \frac{\Gamma\left(\frac{1}{2}\right)}{2^{2z-1}\Gamma(z)},
\end{equation}
we expand the fractions
\begin{eqnarray}
&& \sum\limits_{n=0}^{\infty}{\frac{\left(\frac{Fx_0}{2T}\right)^n}{n!}\frac{\Gamma\left(\frac{n+2}{2}\right)}{\Gamma\left(\frac{n+2}{2}-k\right)}} \nonumber \\
&=&\sum\limits_{m=0}^{\infty}{\left(\frac{Fx_0}{4T}\right)^{2m}\frac{1}{\Gamma\left(m+1-k\right)}\frac{m!\Gamma(1-k)}{m!\Gamma(1-k)}\frac{\Gamma\left(\frac{1}{2}\right)}{\Gamma\left(m+\frac{1}{2}\right)}} \nonumber \\
&&- \frac{Fx_0}{4T}\sum\limits_{m=0}{\left(\frac{Fx_0}{4T}\right)^{2m}\frac{1}{\left(m+\frac{3}{2}-k\right)}\frac{\Gamma\left(\frac{3}{2}-k\right)}{\Gamma\left(\frac{3}{2}-k\right)}\frac{\Gamma\left(\frac{1}{2}\right)}{m!}}.
 \end{eqnarray}
Finally, introducing Pochhammer symbols $(x)_{(n)} = \Gamma(x+n)/\Gamma(x)$, we find
\begin{eqnarray}
&& \sum\limits_{n=0}^{\infty}{\frac{\left(\frac{Fx_0}{2T}\right)^n}{n!}\frac{\Gamma\left(\frac{n+2}{2}\right)}{\Gamma\left(\frac{n+2}{2}-k\right)}} \nonumber \\
 &\hspace*{-1.2cm}=&\hspace*{-0.3cm}\frac{\Gamma\left(1/2\right)}{\Gamma\left(1/2\right)\Gamma(1-k)}\sum\limits_{m=0}^{\infty}{\left(\frac{Fx_0}{4T}\right)^{2m}\frac{1}{m!}\frac{(1)_{(m)}}{(1-k)_{(m)}\left(1/2\right)_{(m)}}} \nonumber \\
 &&\hspace*{-0.3cm}- \frac{Fx_0}{4T}\frac{\Gamma\left(1/2\right)}{\Gamma\left(3/2-k\right)}\sum\limits_{m=0}{\left(\frac{Fx_0}{4T}\right)^{2m}\frac{1}{m!}\frac{1}{\left(3/2-k\right)_{m}}} \nonumber \\
 &\hspace*{-1.2cm}=&\hspace*{-0.3cm}\Gamma\left(\frac{1}{2}\right)\left[{}_1\hat{F}_2\left(1;\frac{1}{2},1-k;\left(\frac{Fx_0}{4T}\right)^2\right) - \frac{Fx_0}{4T}{}_0\hat{F}_1\left(-;\frac{3}{2}-k;\left(\frac{Fx_0}{4T}\right)^2\right)\right]
\end{eqnarray}
in which the ${}_p\hat{F}_q$ are regularised hypergeometric functions \cite{bateman1953higher}. The ageing FPTD reads
 \begin{eqnarray}
 \hspace*{-2.4cm}\wp(t,t_a) &\hspace*{-1.cm}=& \hspace*{-0.4cm}\wp(t+t_a) + \exp{\left(\frac{Fx_0}{2T}\right)}  \sum^{\infty}_{m,n}\frac{t_a^{n+\alpha}}{\Gamma(1+n+\alpha)}\frac{t^{-1-n-\alpha(b m+1)}}{\Gamma\left(-n-\alpha[b m+1]\right)}\left(\frac{4T^2}{F^2K_{\alpha}}\right)^m\frac{\sqrt{\pi}}{m!} \nonumber \\
 &&\hspace*{-1.8cm}\times  \left[{}_1\hat{F}_2 \left(1;\frac{1}{2},1-m; \left(\frac{Fx_0}{4T}\right)^2\right) -\frac{Fx_0}{4T}{}_0\hat{F}_1 \left(-;\frac{3}{2}-m;\left(\frac{Fx_0}{4T}\right)^2 \right)\right]\label{theoforce}
\end{eqnarray}
with the long time limits (cf. \cite{krusemann2014first})
\begin{equation}\label{powerforce}
 \wp(t_a,t) = \left\{
 \begin{array}{ll}
 \displaystyle \frac{\sin{(\pi\alpha)}}{\pi}t_a^{\alpha -1} t^{-\alpha},&t_a \gg t\\[0.3cm]
  \displaystyle \left[\frac{\alpha Tx_0}{FK_{\alpha}\Gamma(1-\alpha)}+\frac{\sin{(\pi\alpha)}}{\pi}t_a^{\alpha}\right]t^{-1-\alpha},&t_a \ll t
 \end{array}
 \right.
\end{equation}
We again find an offset for $t_a\ll t$, which separates the ageing and the non-ageing curve. In figure \ref{forcestuff} we simulated a biased first passage process for different anomalous exponents. We find that the power law behaviour in the strong and weak ageing regions agree well with our long time approximations (\ref{powerforce}).
For $1<\alpha<2$, we follow the arguments in section \ref{boxsec} and reference \cite{margolin2000first}: the FPTD is dominated by the forward waiting time density (\ref{FWT2}) for strong ageing and decays with a finite mean, and almost exponentially, for weak ageing.

\section{Ageing first passage density in quenched and annealed energy landscapes}\label{ageqel}

\begin{figure}
\centering
\includegraphics[width=.75\textwidth]{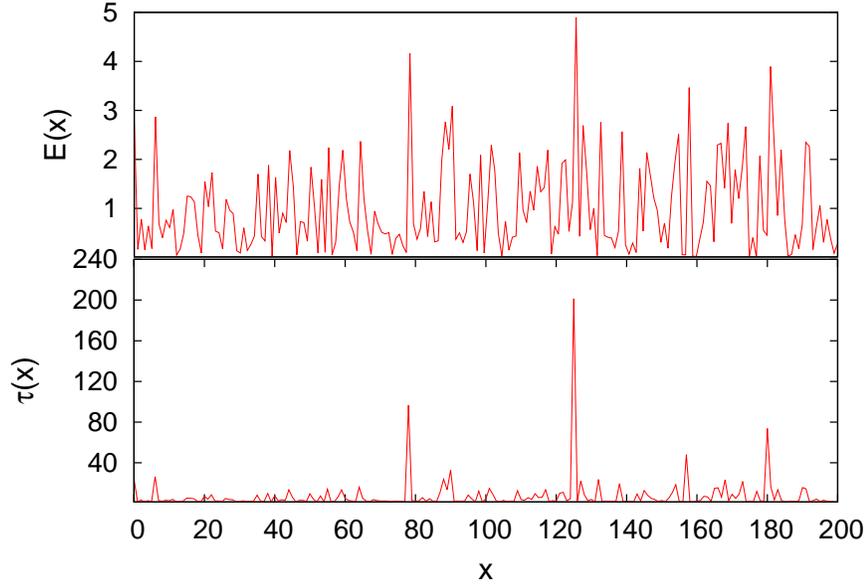}
\caption{Realisation of an energy landscape with $T_G=1$, $T=2/3$ (top).
Bottom: corresponding waiting times $\tau(x)=\exp{\left(E(x)/T\right)}$.
\label{qelwt}}
\end{figure}

\begin{figure}
\centering
\includegraphics[width=.9\textwidth]{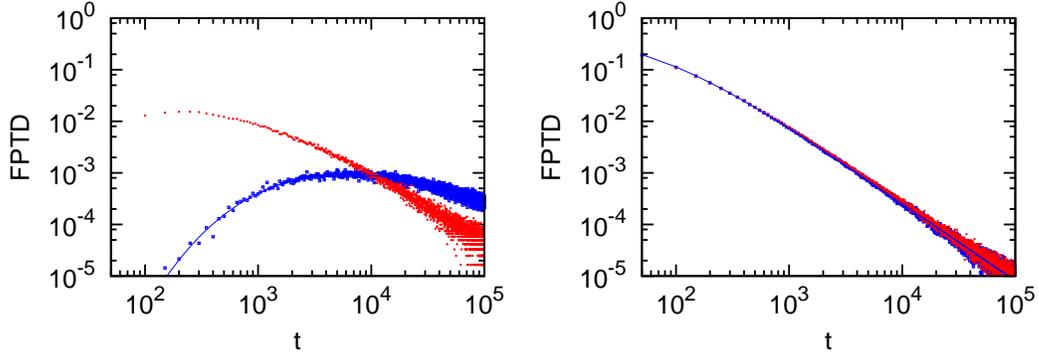}
\caption{FPTD for the quenched energy model in the semi-infinite domain: results
of biased random walk simulations in an annealed (blue) and a quenched landscape
(red) with different results for small bias $F=10^{-5}$ (left) and converging
results for large bias $F=100$ (right). The parameters are $\alpha=1/2$, $t_a=100$,
$T=1/2$, $g=10^{-2}$, $N=200$, and $C=10^{-4}$. The data were averaged over $\sim2
\times10^4$ (left) and $sim 10^6$ (right) runs. The straight blue line represents
the theoretical result for the CTRW.
\label{biasPic}}
\end{figure}

We now consider random walks in a random energy landscape \cite{bertin2003subdiffusion,monthus1996models,bouchaud1992weak,burov2007quenched} on a lattice. Each time the walker is updated it jumps to the left or to the right with equal probability in the unbiased case. At the new site, it is trapped with random energy $E$, which is drawn from an exponential distribution $\rho(E) = T_g^{-1} \exp{\left(-E/T_g\right)}$ \cite{bouchaud1992weak}. The thermal (Kramers) escape from the energy traps dominates the particle motion; it yields a relation between the waiting time $\tau$ and the trap energy $E$, which is given by the Arrhenius law \cite{bouchaud1990anomalous}
\begin{equation}
 \tau = C \exp{\left(\frac{E}{T}\right)}
\end{equation}
where $C$ is the inverse frequency of escape attempts. The distribution for $\tau$ is thus a power law
\begin{equation}
 \psi(\tau) = \alpha C^{\alpha} \tau^{-1-\alpha},
\end{equation}
which yields subdiffusive behaviour, if $\alpha=T/T_G < 1$.
The anomalous diffusion constant is
\begin{equation}
K_{\alpha} = \frac{g^2}{2\Gamma(1-\alpha)C^{\alpha}},
\end{equation}
introducing here the lattice spacing $g$. If a constant bias is to be introduced, we just change the probability to jump to the left from one half to \cite{bertin2003subdiffusion}
\begin{equation}
 p_l = \frac{1}{1 + \exp{(Fg/T)}}.
\end{equation}
We can have two different classes of energy landscapes. In the \emph{annealed random landscape}, in every step a new energy is drawn from the exponential distribution. On the return to a previously visited site, the walker does not find the same energy as before. This case models exactly the subdiffusive CTRW and one finds the subordination relation \cite{metzler2004restaur,barkai2001fractional,burov2012subord,fogedby1994langevin,magdziarz2008equivalence}
\begin{equation}
 n \sim t^{\alpha}
\end{equation}
between the number of jumps and the process time. This is the CTRW process considered in sections \ref{nobiassec} and \ref{biassec}.
In the case of a \emph{quenched energy landscape}, where the energy of each site is fixed (correlations are introduced), the subordination is weakly broken \cite{bouchaud1992weak} and one finds \cite{burov2012subord}
\begin{equation}\label{weaksubbreak}
 n \sim t^{2\alpha /(1+\alpha)}.
\end{equation}
The change in behaviour is due to the multiple appearance of the same waiting times
on revisiting specific locations. A realisation of a quenched energy landscape is
depicted in figure \ref{qelwt}. If the correlations in finding the same waiting
times repeatedly at the same position are sufficiently reduced, for instance by a
large bias or in higher dimensions, one finds the results converging to the
annealed, renewal CTRW. We illustrate the change in the behaviour for CTRW and
motion on a quenched landscape for different bias strength $F$ in figure
\ref{biasPic}: on the left one can see two different curves for small $F$, while
coinciding results on the right appear for large $F$.

\begin{figure}
\centering
\includegraphics[width=.9\textwidth]{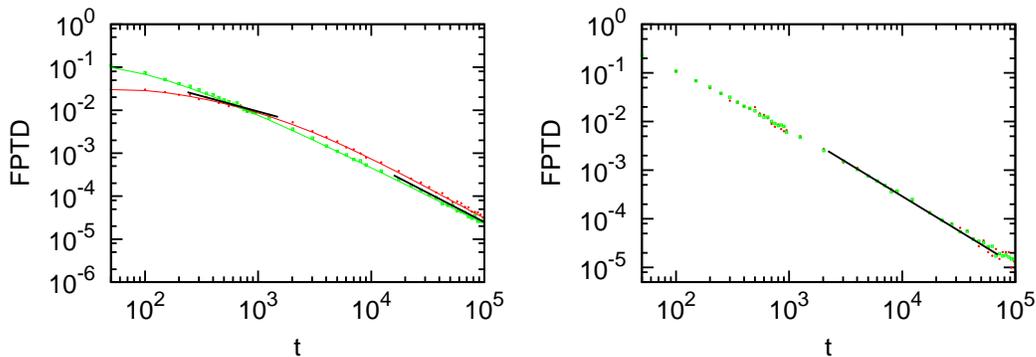}
\caption{FPTD in the semi infinite domain without external bias: results of
annealed landscape simulations (left) and quenched landscape simulations (right)
with (red dots) and without (green squares) ageing. The parameters are $\alpha=
5/7$, $t_a=2\times 10^3$, $T=5/7$, $g=5\times 10^{-2}$, $N=50$ and
$C=10^{-3}$, and averaging is taken over $\sim 2\times 10^4$ runs. The straight
lines are the theoretical results (\ref{fptsemi}), the short black lines indicate
the respective power laws (\ref{3scale}) and (\ref{qpwrlw}). The intermediate regime
in the annealed landscape is too small to see the corresponding power law.
\label{agevsnoage}}
\end{figure}

For an explanation of the long time behaviour of the FPTD in a quenched energy
landscape we use a result from the theory of Brownian random walks, according to
which the number of jumps $n$ is proportional to the time $t$. The asymptotic
behaviour of the Brownian survival probability in the semi-infinite domain is
$\mathscr{S}(t)\sim t^{-1/2}$, and thus we obtain
\begin{equation}
\mathscr{S}(n)\sim n^{-1/2}.
\end{equation}
Using equation (\ref{weaksubbreak}), we expect for the quenched energy landscape
that
\begin{equation}
\label{qpwrlw}
\wp(t)=-\frac{d}{dt}\mathscr{S}(t)\simeq t^{-(1+2\alpha)/(1+\alpha)}.
\end{equation}
One can clearly see the predicted power law in our simulations in figure
\ref{agevsnoage} (right). Interestingly, the unbiased random walk in a quenched
energy landscape exhibits no detectable ageing, unlike the random walk in a
completely random energy landscape (figure \ref{agevsnoage}, graph on the left).
This is due to the fact that the quenched energy landscape features only a
comparatively small sample of random variables relative to the annealed landscape
or a continuous time random walk drawn from the power law waiting time distribution
(\ref{pwrlw}). Ageing enters the system because of immobilisation of the walker for
extremely long waiting times. Such extreme events are found in the large sample of
random waiting times in a CTRW or in an annealed random landscape, but much fewer
realisations occur in the finite QEL. These ageing QEL trajectories are
statistically negligible, as reflected in the FPTD observed here.

\section{Discussion}
We studied the first passage time distribution of an ageing CTRW in three important scenarios, namely the semi-infinite case with and without external bias, and the confined case. We presented the relevant details of the derivation of the exact form of the first passage time densities and extended our long time approximations from Ref. \cite{krusemann2014first} to the asymmetric confined case and to the range $1<\alpha<2$ of the waiting time scaling exponent. In the second part of the paper we show that a different scaling regime appears in a quenched energy landscape, which converges to the CTRW regime with an increasing bias towards the absorbing boundary.
The efficiency of the first passage is generally reduced by ageing. This effect is not only found in subdiffusive processes but is shown to exist even for waiting times of the form (\ref{pwrlw}) with $1<\alpha<3/2$. Even though the corresponding CTRW shows normal behaviour, we can find an aged scaling regime in the strongly aged FPTD, which persists for many decades. For $3/2 < \alpha < 2$, we find no ageing in the unbiased semi-infinite domain, but ageing exists in the bounded domain as well as for biased diffusion.
The ageing effects are so strong, that even a short ageing time is enough to overcome the effects of asymmetry in a confined domain.

The starting position of a particle in a box becomes (almost) irrelevant in the observable regimes.
We find, that the process is accelerated in a quenched energy landscape, which counters the slowing down of ageing. In fact, ageing cannot be observed in a quenched energy landscapes without bias due to the finite sample of energy wells. The slope in the long time limit changes to $-1-\alpha/(\alpha+1)$, but if a bias is introduced, the correlations are overcome and the CTRW regime is recovered excellently.

First passage times are an important quantity for the evaluation of experiments and simulations, as they can be easily obtained and can also be used as diagnostic tools of stochastic processes \cite{condamin2008probing}. Ageing is especially important in biophysics, condensed matter physics and geophysics exhibiting CTRW behaviour, where it corresponds to an immobilisation of particles in crowded cells, changing characteristics of electrical currents in amorphous semiconductors or decreased chemical release rates in groundwater aquifers. We extended the existing results on ageing FPTDs and examined the behaviour in quenched energy landscapes, which are a realistic model for many 1D problems.
Future work could deal with the problem of first passage in higher dimensions (crossing a plane or barrier). Also, it would be interesting to investigate the transition from biased non-ageing quenched energy landscapes to ageing CTRW motion with increasing bias.
Another interesting related topic, the ageing Scher-Montroll transport, is and was examined theoretically \cite{barkai2003aging} and also in time-of-flight experiments \cite{schubert2013mobility}. The experiments in amorphous semiconductors promise excellent testing conditions for the CTRW model for charge carrier transport, as the age of the material can be reliably controlled. In groundwater systems \cite{Scher2002geostuff,berkowitz2006geostuff} the first passage times can be tracked over several decades, so that crossovers between different scaling regimes could be resolved with good resolution. 
In different systems, our results for the crossover times and scaling laws of different power law regimes could be used to determine the age $t_a$ or the stochastic nature of the underlying process.

\ack

RM acknowledges funding from the Academy of Finland (Suomen Akatemia) within the
Finland Distinguished Professor (FiDiPro) programme.

\end{document}